\documentclass[sigconf]{acmart}
\usepackage{algorithm}
\usepackage[noend]{algpseudocode}
\usepackage{csquotes}

\newcommand{\approach}{$TranS^{3}$}

\setcopyright{none}
\newcommand{\response}{24}
\newcommand{\rights}{11}
\newcommand{\explain}{5}

\renewcommand\footnotetextcopyrightpermission[1]{}

\usepackage{fancyhdr}
\AtBeginDocument{%
  \providecommand\BibTeX{{%
    \normalfont B\kern-0.5em{\scshape i\kern-0.25em b}\kern-0.8em\TeX}}}

\begin{document}


\title{\approach{}: A Transformer-based Framework for Unifying Code Summarization and Code Search}
\title[\approach{}: A Transformer-based Framework for Unifying Code Summarization and Search]{\approach{}: A Transformer-based Framework for Unifying Code Summarization and Code Search}


\author{Wenhua Wang}
\email{11760006@mail.sustech.edu.cn}
\affiliation{%
	\institution{Southern University of Science and Technology}
	\streetaddress{P.O. Box 1212}
	\city{Shenzhen}
	\state{Guangzhou}
	\country{China}
	\postcode{43017-6221}
}
\author{Yuqun Zhang}
\email{zhangyq@sustech.edu.cn}
\affiliation{%
  \institution{Southern University of Science and Technology}
  \streetaddress{P.O. Box 1212}
  \city{Shenzhen}
  \state{Guangzhou}
  \country{China}
  \postcode{43017-6221}
}

\author{Zhengran Zeng}
\email{11612527@mail.sustech.edu.cn}
\affiliation{%
	\institution{Southern University of Science and Technology}
	\streetaddress{P.O. Box 1212}
	\city{Shenzhen}
	\state{Guangzhou}
	\country{China}
	\postcode{43017-6221}
}

\author{Guandong Xu}
\email{Guandong.Xu@uts.edu.au}
\affiliation{%
	\institution{University of Technology, Sydney}
	\city{Sydney}
	\state{NSW}
	\country{Australia}
	\postcode{43017-6221}
}


\begin{abstract}
	Code summarization and code search have been widely adopted in software development and maintenance. However, few studies have explored the efficacy of unifying them. 
	In this paper, we propose \approach{}, a transformer-based framework to integrate code summarization with code search. Specifically, for code summarization, \approach{} enables an actor-critic network, where in the actor network, we encode the collected code snippets via transformer- and tree-transformer-based encoder and decode the given code snippet to generate its comment. Meanwhile, we iteratively tune the actor network via the feedback from the critic network for enhancing the quality of the generated comments. Furthermore, we import the generated comments to code search for enhancing its accuracy. To evaluate the effectiveness of \approach{}, we conduct a set of experimental studies and case studies where the experimental results suggest that \approach{} can significantly outperform multiple state-of-the-art approaches in both code summarization and code search and the study results further strengthen the efficacy of \approach{} from the developers' points of view. 
  
\end{abstract}

\maketitle

\section{Introduction}\label{sec:introduction}
Code summarization and code search have become increasingly popular in software development and maintenance \cite{leclair2019neural, movshovitz2013natural, sridhara2010towards,lv2015codehow, gu2018deep, yao2019coacor}, because they can help developers understand and reuse billions of lines of code from online open-source repositories and thus significantly enhance software development and maintenance process \cite{yao2019coacor}. In particular, since much of the software maintenance effort is spent on understanding the maintenance task and related software source code \cite{lientz1980software}, effective and efficient documentation is quite essential to provide high-level descriptions of program tasks for software maintenance. To this end, code summarization aims to automatically generate natural language comments for documenting code snippets \cite{moreno2017automatic}. On the other hand, over years various open-source and industrial software systems have been rapidly developed where the source code of these systems is typically stored in source code repositories. Such source code can be treated as important reusable assets for developers because they can help developers understand how others addressed similar problems for completing their program tasks, e.g., testing~\cite{deeproad,mutationtesting,regressiontesting, simulee1, deepbillboard1, iottesting, simulee2}, fault localization~\cite{deepfl,prfl,prfl1}, program repair and synthesis~\cite{bytecode,history, sttt1, aucs}, in multiple software development domains~\cite{smartvm, mssurvey,bigvm}. Correspondingly, there also raises a strong demand for an efficient search process through a large codebase to find relevant code for helping programming tasks. To this end, code search refers to automatically retrieving relevant code snippets from a large code corpus given natural language queries. 

The recent research progress towards code summarization and code search can mainly be categorized to two classes: \textit{information-retrieval}-based approaches and \textit{deep-learning}-based approaches. To be specific, the \textit{information-retrieval}-based approaches derive the natural language clues from source code, 
compute and rank the similarity scores between them and source code/natural language queries for recommending comments/search results \cite{wong2013autocomment, movshovitz2013natural, lu2015query, lv2015codehow}. The \textit{deep-learning}-based approaches use deep neural networks to encode source code/natural language into a hidden space, and utilize neural machine translation models for generating code comments and computing similarity distance to derive search results \cite{hu2018summarizing, chen2018neural, gu2018deep, akbar2019scor, yao2019coacor}. 

Based on the respective development of code summarization and code search techniques, we infer that developing a unified technique for optimizing both domains simultaneously is not only mutually beneficial but also feasible. In particular, on one hand, since the natural-language-based code comments can reflect program semantics to strengthen the understanding of the programs~\cite{leclair2019neural}, adopting them in code search can improve the matching process with natural language queries~\cite{yao2019coacor}. Accordingly, injecting code comments for code search is expected to enhance the search results~\cite{Scholer2014Query}. On the other hand, the returned search results can be utilized as an indicator of the accuracy of the generated code comments to guide their optimization process. Moreover, since code summarization and code search can share the same technical basis as mentioned above, it can be inferred that it is feasible to build a framework to unify and advance both the domains. 
Therefore, it is essential to integrate code summarization with code search.

Although integrating code summarization with code search can be promising, there remains the following challenges that may compromise its performance: (1) state-of-the-art code summarization techniques render inferior accuracy. According to the recent advances in code summarization \cite{wong2013autocomment, iyer2016summarizing, hu2018deep}, the accuracy of the generated code comments appears to be inferior for real-world applicability (around 20\% in BLEU-1 with many well-recognized benchmarks). Integrating such code comments might lead to inaccuracies of matching natural language queries and further compromise the performance of code search. (2) how to effectively and efficiently integrate code summarization with code search remains challenging. Ideally, the goal of integrating code summarization with code search is to optimize the performance of both domains rather than causing trade-offs. Moreover, such integration is expected to introduce minimum overhead. To this end, it is essential to propose an effective and efficient integration approach.

To tackle the aforementioned problems, in this paper, we propose a framework, namely \approach{} for optimizing both code summarization and code search based on a recent NLP technique---transformer \cite{vaswani2017attention}. Unlike the traditional CNN-based approaches that suffer from long-distance dependency problem \cite{wang2016dimensional} and RNN-based approaches that suffer from excessive load imposed by sequential computation \cite{huang2013accelerating}, transformer advances in applying the self-attention mechanism which can parallelize the computation and preserve the integral textual weights for encoding to achieve the optimal accuracy of text representation \cite{vaswani2017attention}. 

\approach{} consists of two components: the code summarization component and code search component. 
Specifically, the code summarization component is initialized by preparing a large-scale corpus of annotated $ <code; comment> $ pairs to record all the code snippets with their corresponding comments as the training data. Next, we extract the semantic granularity of the training programs for constructing a tree-transformer to encode the source code into hidden vectors.
Furthermore, such annotated pair vectors are injected into our deep reinforcement learning model, i.e., the actor-critic framework, for the training process, where the actor network is a formal encoder-decoder model to generate comments given the input code snippets; and the critic network evaluates the accuracy of the generated comments according to the ground truth (the input comments) and give feedback to the actor network. At last, given the resulting trained actor network and a code snippet, its corresponding comment can be generated. 
Given a natural language query, the code search component is launched by encoding the natural language query, the generated code comments, and the code snippets into the vectors respectively via transformer and tree-transformer. Next, we compute similarity scores between query/code vectors and query/comment vectors for deriving and optimizing their weighted $scores$. 
Eventually, we rank all the code snippets according to their $score$s for recommending the search results. The underlying transformer in \approach{} can enhance the quality of the generated code and  thus strengthen the code search results by importing the impact from the generated comments. Moreover, since the code search component applies the encoder trained by the code summarization component without incurring extra training process, its computing overhead can be maintained minimum.
	
	To evaluate the effectiveness and efficiency of \approach{}, we conduct a set of experiments based on the GitHub dataset in \cite{Barone2017A} which includes over 120,000 code snippets of Python functions and their corresponding comments. The experimental results suggest that \approach{} can outperform multiple state-of-the-art approaches in both code summarization and code search, e.g., \approach{} can significantly improve the code summarization accuracy from 47.2\% to 141.6\% in terms of BLEU-1  and the code search accuracy from 5.1\% to 28.8\% in terms of MRR compared with the selected state-of-the-art approaches.
	In addition, we also conduct case studies for both code summarization and code search where the study results further verify the effectiveness of \approach{}.

In summary, the main contributions of this paper are listed as follows:
	\begin{itemize}
		\item \textbf{Idea.} To the best of our knowledge, we build the first transformer-based framework for integrating code summarization and code search, namely \approach{}, that can optimize the accuracy of both domains. 
		
		\item \textbf{Technique.} To precisely represent the source code, we design a transformer-based encoder and a tree-transformer-based encoder for encoding code and comments by injecting the impact from the semantic granularity of well-formed programs. 

		\item \textbf{Evaluation.} To evaluate \approach{}, we conduct a substantial number of experiments based on real-world benchmarks. The experimental results suggest that \approach{} can outperform several existing approaches in terms of accuracy of both code summarization and code search. In addition, we also conduct empirical studies with developers. The results suggest that the quality of the generated comments and search results are widely acknowledged by developers. 

	\end{itemize}

	The reminder of this paper is organized as follows. Section \ref{sec:background} illustrates some preliminary background techniques. Section \ref{sec:example} gives an example to illustrate our motivation for unifying code summarization and code search. Section \ref{sec:approach} elaborates the details of our proposed approach. Section \ref{sec:experiment} demonstrates the experimental and study results and analysis. Section \ref{sec:threats} introduces the threats to validity. Section \ref{sec:relatedwork} reviews the related work. Section \ref{sec:conclusion} concludes this paper.

\section{Background}\label{sec:background}
    In this section, we present the preliminary background techniques relevant to \approach{}, including language model, transformer, and reinforcement learning, which are initialized by introducing basic notations and terminologies. Let $\mathbf{x}=(x_1, x_2,\ldots, x_{|\mathbf{x}|})$ denote the code sequence of one function, where $ x_t$ represents a token of the code, e.g., ... ``\textit{def}'', ``\textit{fact}'', or ``\textit{i}'' in a Python statement ``\textit{def fact(i)}:''. 
	Let $\mathbf{y}=(y_1, y_2,\ldots, y_{|\mathbf{y}|})$ denote the sequence of the generated comments, where {\scriptsize $|\mathbf{y}|$} denotes the sequence length.  
	Let $T$ denote the maximum step of decoding in the encoder-decoder framework. We use notation $y_{l \ldots m}$ to represent the comment subsequence $y_l, \ldots , y_m$ and $\mathcal{D}=\{(\mathbf{x}_{N},\mathbf{y}_{N})\}$ as the training dataset, where $N$ is the size of training set. 

	\subsection{Language Model}
	A language model refers to the decoder of neural machine translation which is usually constructed as the probability distribution over a particular sequence of words.
	Assuming such sequence with its length $ T $, the language model defines $p(y_{1:T})$ as its occurrence probability which is usually computed based on the conditional probability from a window of $n$ predecessor words, known as $n$-gram \cite{wang2016bugram}, as shown in Equation \ref{conditionprobability}. 
		
	\begin{equation}
	p(y_{1:T})=\prod_{t=1}^{i=T}p(y_t|y_{1:t-1})\approx \prod_{t=1}^{t=T}p(y_t|y_{t-(n-1):t-1})
	\label{conditionprobability}
	\end{equation}
		
	While the $n$-gram model can only predict a word based on a fixed number of predecessor words, a neural language model can use predecessor words with longer distance to predict a word based on deep neural networks which include three layers: an input layer which maps each word $ x_t $ to a vector, a recurrent hidden layer which recurrently computes and updates a hidden state $ h_t $ after reading $ x_t $, and an output layer which estimates the probabilities of the subsequent words given the current hidden state. 
	In particular, the neural network reads individual words from the input sentence, and predicts the subsequent word in turn. For the word $y_{t}$, the probability of its subsequent word $y_{t+1}$, $p(y_{t+1}|y_{1:t})$ can be computed as in Equation \ref{predictstate}: 

	\begin{equation}
	p(y_{t+1}|y_{1:t}) = g(\mathbf{h}_t)
	\label{predictstate}
	\end{equation}
	where $g$ is a stochastic output layer (e.g., a softmax for discrete outputs) that generates output tokens with the hidden state $\mathbf{h}_t$ computed as Equation \ref{hiddenstate}:
	
	\begin{equation}
	\mathbf{h}_t = f(\mathbf{h}_{t-1}, w(x_t))
	\label{hiddenstate}
	\end{equation}
	
	where $w(x_t)$ denotes the weight of the token $x_t$. 
	
\subsection{Transformer} 
Many neural machine translation approaches integrate the attention mechanism with sequence transduction models for enhancing the accuracy. However, the encoding networks are still exposed with challenges. To be specific, the CNN-based encoding networks are subjected to long-distance dependency issues and the RNN-based encoding networks are subjected to the long-time computation. 	
To address such issues, transformer \cite{vaswani2017attention} is proposed to effectively and efficiently improve the sequence representation by adopting the self-attention mechanism only. Many transformer-based models e.g., BERT \cite{devlin2018bert}, ERNIE \cite{sun2019ernie}, XLNET \cite{yang2019xlnet}, have been proposed and verified to dramatically enhance the performance of various NLP tasks such as natural language inference , text classification, and retrieval question answering \cite{Bowman2015A, Voorhees2001The}.

\begin{figure}
		\includegraphics[width=0.47\textwidth]{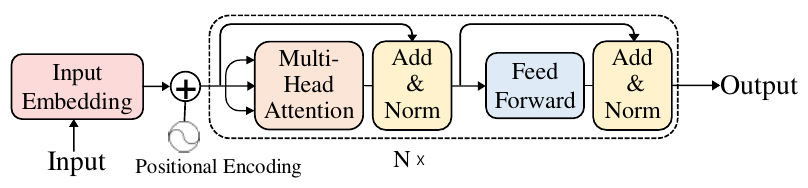}
		\caption{The Transformer Model Architecture.} 
		\label{fig:transformer}
	\end{figure}

Transformer consists of $ N $ identical layers where each layer consists of two sub-layers. The first sub-layer realizes a multi-head self-attention mechanism, and the second sub-layer is a simple, position-wise fully connected feed-forward neural network, as shown in Figure \ref{fig:transformer}. Note that the output of the first sub-layer is input to the second sub-layer and the outputs of both the sub-layers need to be normalized prior to the subsequent process.
	
	\subsubsection{Self-attention Mechanism}
	
	The attention function can be described as mapping a query and a set of key-value pairs to an output, where the query, keys, values, and output are all vectors. The output is computed as a weighted sum of the values, where the weight assigned to each value is computed by a compatibility function of the query with the corresponding key.
	 
	The input consists of queries, keys and values of the dimension $ d_k $. Accordingly, transformer computes the dot products of the query with all keys, divides each resulting element by $ \sqrt{d_k}$, and applies a softmax function to obtain the weights on the values. 
	In practice, we simultaneously compute the attention function on a set of queries which are packed together into a matrix $ Q $. In addition, the keys and values are also packed together into matrices $ K $ and $ V $. Therefore, the matrix of outputs can be computed as:
	
	\begin{equation}
	Attention(Q,K,V) = softmax(\dfrac{QK^{T}}{\sqrt{d_k}})V
	\label{attention}
	\end{equation}
	
	Instead of implementing a single attention function, transformer adopts a multi-head attention which allows the model to jointly attend to information from different representation subspaces at different positions. 
		
	The self-attention mechanism derives the relationships between the current input token and all the other tokens to determine the current token vector for the final input representation. 
	By taking advantage of the overall token weights, such mechanism can dramatically alleviate the long-distance dependency problem caused by the CNN-based transduction models, i.e., compromising the contributions of the long-distance tokens.
	Moreover, the multi-head self-attention mechanism can parallelize the computation and thus resolve the excessive computing overhead caused by the RNN-based transduction models which sequentially encode the input tokens.

	\subsubsection{Position-wise Feed-Forward Neural Network}
	In addition to multi-head self-attention sub-layers, each of the layers contains a fully connected feed-forward neural network, which is applied to each position separately. Since transformer contains no recurrence or convolution, in order to utilize the order of the sequence, transformer injects ``positional encodings'' to the input embedding.
 
    Since transformer has been verified to be dramatically effective and efficient in encoding word sequences, we infer that by representing code as a sequence, transformer can also be expected to excel in the encoding efficacy. Therefore, in \approach{}, we adopt transformer as the encoder. 
    
\subsection{Reinforcement Learning for Code Summarization}
	In code summarization, reinforcement learning (RL)\cite{Thrun2005Reinforcement} refers to interacting with the ground truth, learning the optimal policy from the reward signals, and generating texts in the testing phase. 
	It can potentially solve the exposure bias problem introduced by the maximum likelihood approaches which is used to train the RNN model. Specifically in the inference stage, a typical RNN model generates a sequence iteratively and predicts next token conditioned on its previously predicted ones that may never be observed in the training data \cite{yu2017seqgan}. Such a discrepancy between training and inference can become cumulative along with the sequence and thus prominent as the length of sequence increases. While in the reinforcement-learning-based framework, the reward, other than the probability of the generated sequence, is calculated to give feedback to train the model to alleviate such exposure bias problem.  
	Such text generation process can be viewed as a Markov Decision Process (MDP) $\{state, action, policy, reward\}$. Specifically in the MDP settings, $state$ $\mathbf{s}_t$ at time $t$ consists of the code snippets $\mathbf{x}$ and the predicted words ${y_0,y_1,\ldots,y_t}$. The $action$ space is defined as the dictionary $\mathcal{Y}$ where the words are drawn, i.e., $y_t \subset \mathcal{Y}$. Correspondingly, the $state$ transition function $P$ is defined as $\mathbf{s}_{t+1} = \{\mathbf{s}_t, y_{t}\}$, where the $action$ (word) $y_{t}$ becomes a part of the subsequent $state$ $\mathbf{s}_{t+1}$ and the $reward$ $r_{t+1}$ can be derived. 
	The objective of the generation process is to find a $policy$ that iteratively maximizes the expected $reward$ of the generated sentence sampled from the model's $policy$, as shown in Equation \ref{rnncode},
	
	\begin{equation}
	\underset{\theta}{\max}\mathcal{L}(\theta) = \underset{\theta}{\max}\mathbb{E}_{\underset{\hat{\mathbf{y}}\sim P_{\theta}(\cdot|\mathbf{x})}{\mathbf{x}\sim \mathcal{D}}}[R(\hat{\mathbf{y}},\mathbf{x})]
	\label{rnncode}
	\end{equation}
	
	where $\theta$ is the policy parameter to be learned, $\mathcal{D}$ is the training set, $\hat{\mathbf{y}}$ denotes the predicted $action$s/words, and $R$ is the reward function. 
	
	To learn the policy, many approaches have been proposed, which are mainly categorized into two classes \cite{sutton1998introduction}: (1) the policy-based approaches (e.g., Policy gradients \cite{williams1992simple}) which optimize the policy directly via policy gradient and (2) the value-based approaches (e.g., Q-learning \cite{watkins1992q}) which learn the Q-function, and at each time the agent selects the action with the highest Q-value. It has been verified that the policy-based approaches may suffer from a variance issue and the value-based approaches suffer from a bias issue \cite{keneshloo2018deep}. To address such issues, the Actor-Critic learning approach is proposed \cite{Konda2003Actor} to combine the strengths of both policy- and value-based approaches where the actor chooses an action according to the probability of each action and the critic assigns the value to the chosen action for speeding up the learning process for the original policy-based approaches. 
	
	In this paper, we adopt the actor-critic learning model for code summarization of \approach{}. 

\section{Illustrative Example}\label{sec:example}
    
\begin{figure}
		\includegraphics[width=0.48\textwidth]{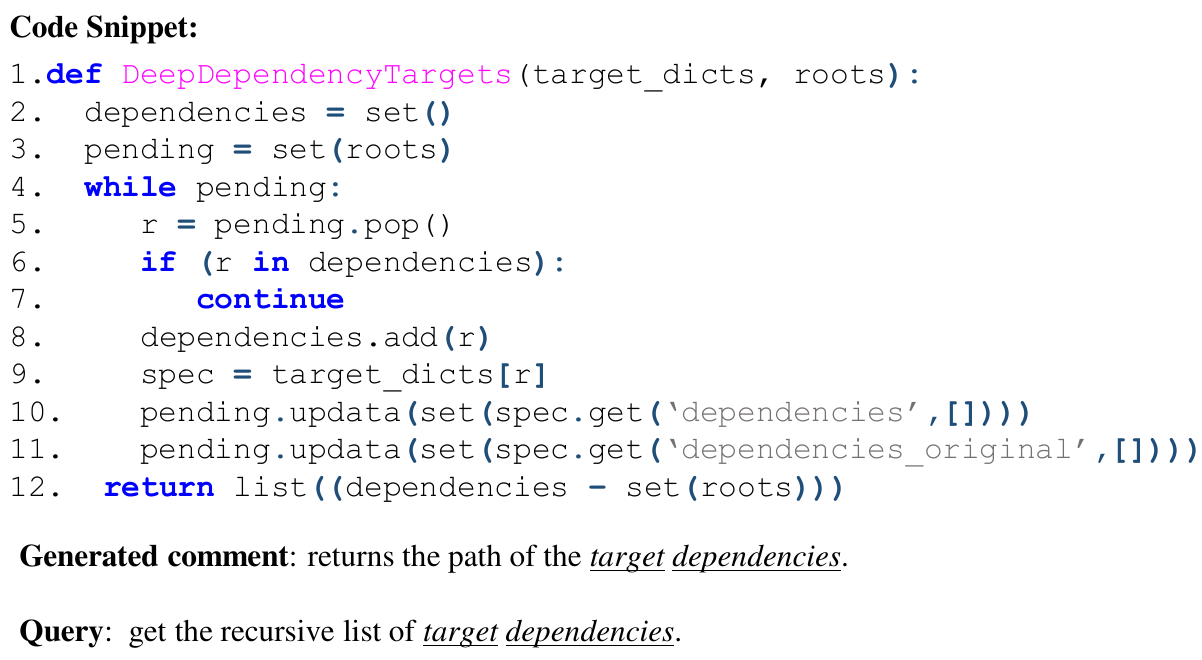}
		\caption{An example Python code snippet and the corresponding query and generated comment.}
		\label{fig:example}
	\end{figure}

	In this section, we use a sample Python code snippet to illustrate our motivation for unifying code summarization and code search. Figure \ref{fig:example} shows the Python code snippet, the comment generated by our approach and its associated natural language query in our dataset.
    Traditional code search approaches usually compute the similarity scores of the query vector and the code snippet vectors for recommending and returning the relevant code snippets. On the other hand, provided the comment information, it is plausible to enhance the code search results by enabling an additional mapping process between the query and the comments corresponding to the code snippets. For example, in Figure \ref{fig:example}, given the query ``\textit{get the recursive list of target dependencies}'', although the code snippet can provide some information such as ``\textit{dependencies}'', ``\textit{target}'', which might be helpful for being recommended, its efficacy can be compromised due to the disturbing information such as ``\textit{dicts}'', ``\textit{set}'', ``\textit{pending}'' in the code snippet. It is expected to enhance the search result by integrating the comment information during the searching process when it has the identical ``\textit{target}'', ``\textit{dependencies}'' with the query.   
    To this end, we infer that a better code search result can be expected if high-quality comment information can be integrated in the code search process.
    
	\begin{figure*}
		\includegraphics[width=0.7\textwidth]{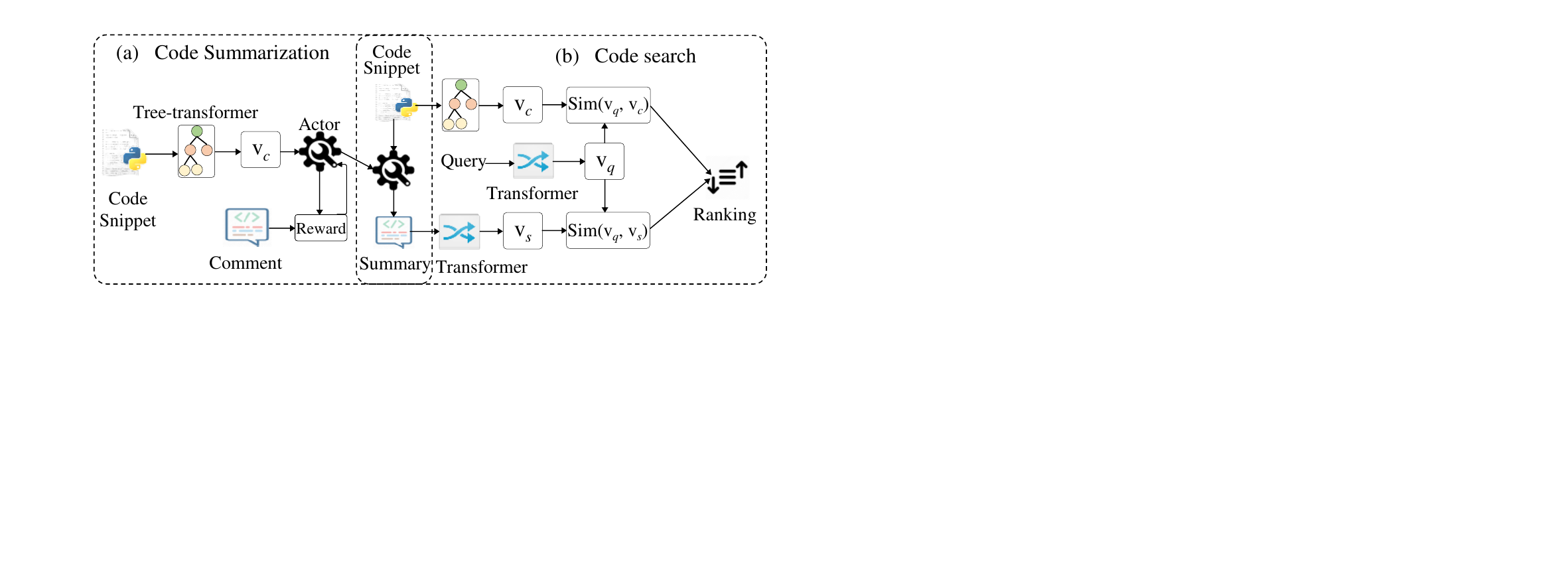}
		\caption{The Overview of Our Proposed approach \approach{}. (a) the code summarization part; (b) the code search part.}
		\label{fig:overview}
	\end{figure*}

\section{The Approach of \approach{}}\label{sec:approach}	

We formulate the research problem of integrating code summarization with code searchas as follows:
	
	\begin{itemize}
		\item First, we attempt to find a policy that generates a sequence of words $\mathbf{y}=(y_1, y_2,\ldots, y_{|\mathbf{y}|})$ from dictionary $\mathcal{Y}$ to annotate the code snippets in the corpus as their comments. Next, given a natural language query $\mathbf{x}=(x_1, x_2,\ldots, x_{|\mathbf{x}|})$, we aim to find the code snippets that can satisfy the query under the assistance of the generated comments.
	\end{itemize}

	\begin{figure}
		\includegraphics[width=0.32\textwidth]{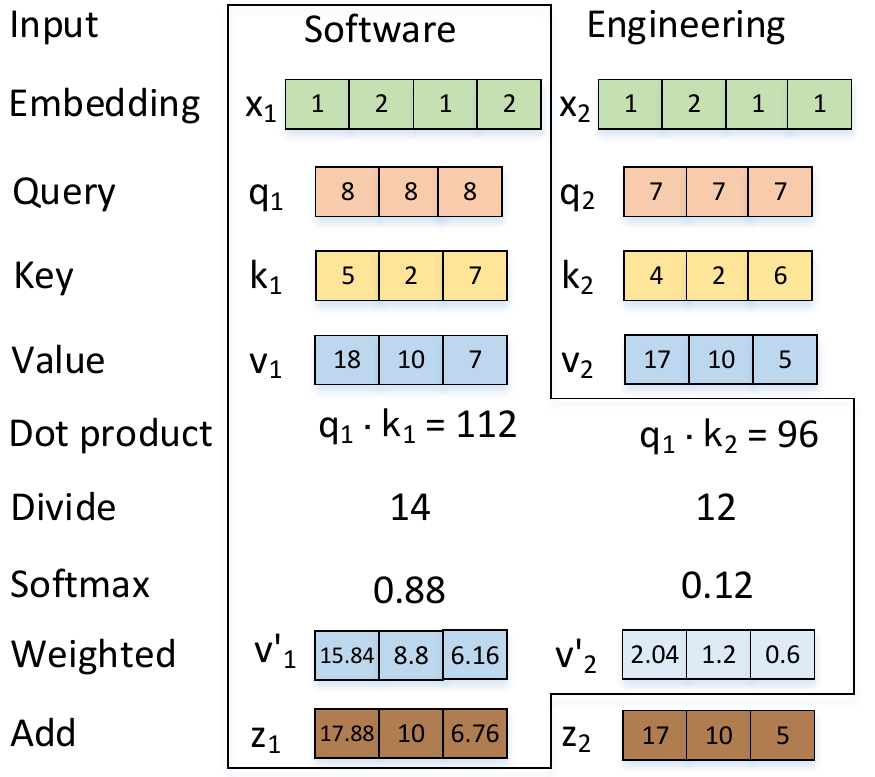}
		\caption{An example of Self-Attention Mechanism.}
		\label{fig:attention}
	\end{figure}
	
To address such research problem, we propose \approach{} with its framework shown in Figure \ref{fig:overview}, where Figure \ref{fig:overview}(a) presents the code summarization part and Figure \ref{fig:overview}(b) presents the code search part.

\subsection{Transformer- and Tree-Transformer-based Encoder} \label{sec:encoder}
In \approach{}, we utilize transformer to build the encoder. Specifically, we develop the transformer-based encoder to encode the comments, the query and each program statement. 
Moreover, we develop a tree-transformer-based encoder that exploits the semantic granularity information of programs for enhancing the program encoding accuracy. 

\textbf{Transformer-based Encoder.}
The transformer-based encoder is initialized by embedding the input tokens into vectors via word embedding \cite{Mikolov2013Efficient}. 
Specifically, we tokenize the natural language comments/queries based on their intervals and the code based on a set of symbols, i.e., {\{ . , " ' : * () ! - \_ (space)\}}. Next, we apply word embedding to derive each token vector in one input sequence. 
Furthermore, for each token vector $x_i$, we derive its representation according to the self-attention mechanism as follows: (1) deriving the query vector $ q_i$, the key vector $k_i$, and the value vector $v_i$ by multiplying $x_i$ with a randomly-generated matrix, (2) computing the scores of $x_i$ from all the input token vectors by the dot product of $ q_i \cdot k_j $, where $j\epsilon [1, n]$ and $n$ denotes the number of input tokens, (3) dividing the scores of $x_i$ by $\sqrt{d_k}$ where $d_k$ denotes the dimension number of $k_i$ and normalizing the results by softmax to obtain the weights (contributions) of all the input token vectors, (4) multiplying such weights and their corresponding value vectors to obtain an interim vector space $v^{'}$, and (5) summing all the vectors in $v^{'}$ for deriving the final vector of $x_i$, $z_i$. As a result, all the token vectors are input to the feed-forward neural network to obtain the final representation vector of the input sequence of natural language comment. 

We use Figure \ref{fig:attention} as an example to illustrate how the transformer-based encoder works, where the token vectors of ``\textit{Software}'' and ``\textit{Engineering}'' are embedded as $x_1$ and $x_2$ respectively. For $x_1$, its corresponding $q_1$, $k_1$, and $v_1$ are derived in the beginning. Next, its scores from all the token vectors, i.e., $x_1$ and $x_2$, can be computed by $q_1 \cdot k_1$ (112) and $q_1 \cdot k_2$ (96). Assuming $d_k$ is 64, by dividing the resulting dot products by $\sqrt{d_k}$ and normalizing, the weights of $x_1$ and $x_2$ can be computed as 0.88 and 0.12. At last, $z_1$ can be derived by 0.88*$v_1$ + 0.12*$v_2$. 

\textbf{Tree-Transformer-based Code Encoder.} 
It can be observed that well-formed source code can reflect the program semantics through its representations, e.g., the indents of Python.
In general, in a well-formed program, the statement with fewer indents tends to indicate more abstracted semantics than the one with longer indents. 
Therefore, we infer that incorporating the indent-based semantic granularity information for encoding can inject program semantics for program comprehension and thus be promising to enhance the encoding accuracy. 
Such injection can potentially be advanced when leveraging transformer. In particular, in addition to the original self-attention mechanism which determines the token vector score by only importing the token-level weights, statement-level impacts can be injected by analyzing statement indents, obtaining the semantic hierarchy of the code, and realizing the hierarchical encoding process. 

	\begin{algorithm}[!t]
		\caption{Tree Transformer Encoding Algorithm.}\label{alg:tree}
		\begin{flushleft}
			\hspace*{\algorithmicindent} \textbf{Input }:  ordered tree ()
		\end{flushleft}
		\begin{flushleft}
			\hspace*{\algorithmicindent} \textbf{Output}: vector representation of the tree 
		\end{flushleft}
		\begin{algorithmic}[1]
			\Function{PostorderTraverse}{}
			\State  node\_list $\leftarrow$ root node
			\If{isLeaf(root\_node)} 
			\State  \textbf{return} Transformer(node\_list); 
			\Else
			\For {$i$ in range(len(root\_node's children))} 
			\State  node\_list.append(PostOrderTraverse($i$'s children))
			\EndFor 
			\State return Transformer(node\_list)
			\EndIf
			\EndFunction
		\end{algorithmic}
	\end{algorithm}

In this paper, we design a tree-transformer-based encoder that incorporates indent-based semantic granularity for encoding programs.
Firstly, we construct an ordered tree according to the indent information of a well-formed program. In particular, by reading the program statements in turn, we initialize the tree by building the root node out of the function definition statement. Next, we iteratively label each of the subsequent statements with an indent index assigned by counting the indents such that the statements with the same indent index $ i $ are constructed as the ordered sibling nodes and the preceding statement above such statement block with the indent index $i-1$ is constructed as their parent node. 
Secondly, we encode each node (i.e., each statement) of the tree into a vector by transformer. At last, we build the tree-transformer accordingly to further encode all the vector nodes of the tree for obtaining the code snippet representations. Specifically, we traverse the tree in a post-order manner. Assuming a node $n_i$ and its parent node $n_j$, if $n_i$ is a leaf node, we replace the vector of $n_j$, namely $V_{n_j}$ by the vector list \{$V_{n_i}$, $V_{n_j}$\} and subsequently traverse $n_j$'s other child nodes; otherwise, we traverse $n_i$'s child nodes. Next, we encode node $n_j$ with the updated vector list \{$V_{n_i}$, $V_{n_j}$\} by transformer when it has no child nodes. The tree-transformer encoding process is shown as Algorithm \ref{alg:tree}.

\begin{figure}
		\includegraphics[width=0.37\textwidth]{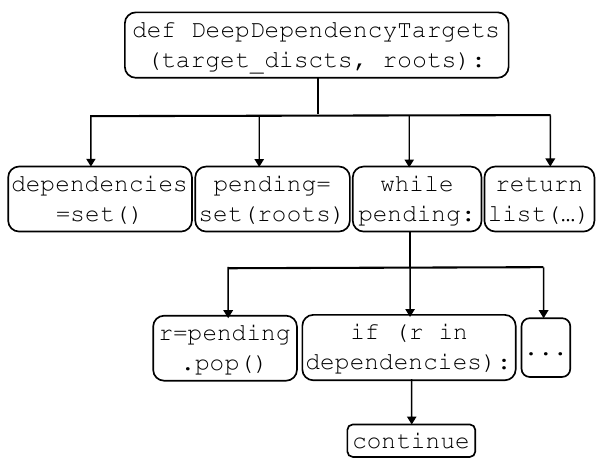}
		\caption{The Source Code and the Tree Structure.}
		\label{fig:tree}
	\end{figure}

Figure \ref{fig:tree} illustrates indent-based tree representation of the code snippet given in Figure \ref{fig:example}. We use this example to describe how the tree-transformer-based encoder works. 
Specifically in Figure \ref{fig:example}, we construct nodes ``Dependencies = set()'', ``pending=set(roots)'', ``while pending:'' and ``return list(...)'' as siblings because they are assigned with the same indents and  one-shorter-indent preceding statement ``def DeepDependencyTargets(target\_dicts, roots):'', which is constructed as their parent node.
Then, we encode all the statement nodes into vectors by transformer respectively. 
Next, as the root's child nodes ``Dependencies = set()'' and ``pending=set(roots)'' are leaf nodes, we replace the root vector by the vector list of them three. Then, since the root's child node ``while pending:'' is not the leaf node,  we first encode its child node ``if (r in dependencies):'' with ``continue'' by transformer, and then encode the resulting vector with the siblings of ``while pending:'' and ``if (r in dependencies):'' together by transformer. At last, we encode the root node with all its child nodes to obtain the final representation of this code snippet.

\subsection{Code Summarization}
 Initialized by collecting code snippets with their associated comments and forming $ <code; comment>$ pairs for training the code summarization model, the code summarization component is implemented via reinforcement learning (i.e., the actor-critic framework), where the actor network establishes an encoder-decoder mechanism to derive code comments and the critic network iteratively provides feedback for tuning the actor network. In particular, 
the actor network leverages a transformer-based or a tree-transformer-based encoder to encode the collected code into hidden space vectors and applies a transformer-based decoder to decode them to natural language comments. 
Next, by computing the similarity between the generated and the ground-truth comments, the critic network iteratively provides feedback for tuning the actor network. As a result, given a code snippet, its corresponding natural language comment can be generated based on the trained code summarization model. 

\subsubsection{Actor Network. } The actor network is composed of an encoder and decoder. 

\textbf{Encoder. }We construct the tree representation of the source code and establish a tree-transformer, described as Section \ref{sec:encoder}, to encode the source code into hidden space vectors for the code representation.
	
\textbf{Decoder.}
After obtaining the code snippet representations, \approach{} implements the decoding process for them, i.e., generating comments from the hidden space, to derive their associated natural language comments. 
	
The decoding process is launched by generating an initial decoding state $s_0 = \{x\}$ by encoding the given code snippet. 
At step $t$, state $s_t$ is generated to maintain the source code snippet and the previously generated words $y_{1...t-1}$, i.e., $s_t = \{x, y_{1...t-1}\}$. Specifically, the previously generated words $y_{1...t-1}$ are encoded into a vector by transformer and subsequently concatenated with state  $s_{t-1}$. Our approach predicts the $t$th word by using a softmax function. Let $p(y_t|\mathbf{s}_t)$ denote the probability distribution of the $t$th word $y_t$ in the state $\mathbf{s}_t$, we can obtain the following equation: 
	\begin{equation}
	p(y_t|\mathbf{s}_t)=softmax(\mathbf{W}_s\mathbf{s}_t+\mathbf{b}_s)
	\end{equation}
	Next, we update $\mathbf{s}_{t}$ to $\mathbf{s}_{t+1}$ to generate the next word. This process is iterated till it exceeds the max-step or generates the end-of-sequence (EOS) token for generating the whole comment corresponding to the code snippet. 
	
	\subsubsection{Critic Network}
	To enhance the accuracy of the generated code comments, 
	\approach{} applies a critic network to approximate the value of the generated comments at time $t$ to issue a feedback to tune the network iteratively.  
	Unlike the actor network which outputs a probability distribution, the critic network outputs a single value on each decoding step. 
	To illustrate, given the generated comments and the reward function  $r$, the value function $V$ is defined to predict the total reward from the state $\mathbf{s}_t$ at time $t$, which is formulated as follows,
	
	\begin{equation}\label{eq:value}
	V(\mathbf{s}_t)=\mathbb{E}_{\overset{\mathbf{s}_{t+1:T},}{y_{t:T}}}\left [\sum_{l=0}^{T-t}r_{t+l}|y_{t+1},\cdots,y_{T}, \mathbf{h} \right]
	\end{equation}
	where $T$ is the max step of decoding and $\mathbf{h}$ is the representation of code snippet. 
	By applying the reward function, we can obtain an evaluation score (e.g., BLEU) when the generation process of the comment sequences is completed. Such process is terminated when the associated step exceeds $T$ or generates the end-of-sequence (EOS) token. For instance, a BLEU-based reward function can be calculated as:

	\begin{equation}\label{eq:bleu}
	r=exp(\frac{1}{N} * \sum_{i=1}^{N}logp_n )
	\end{equation}
	where $p_n=\frac{\sum_{n-gram \in c} count(n-gram)}{\sum_{n-gram\in c^{'}}count(n-gram)}$, and $c$ is the generated comment an $c^{'}$ is the ground truth.

	\subsubsection{Model Training}
For the actor network, the training objective is to minimize the negative expected reward, which is defined as $\mathcal{L}(\theta) =- \mathbb{E}_{y_{1,\ldots,T}\sim \pi}(\sum_{l=t}^{T}r_t)$, where $\theta$ is the parameter set of the actor network. Defining policy as the probability of a generated comment, we adopt the policy gradient approach to optimize the policy directly, which is widely used in reinforcement learning. 

	The critic network attempts to minimize the following loss function, 
	\begin{equation}
		\mathcal{L}(\phi) = \frac{1}{2}\left \| V(\mathbf{s}_t) - V_\phi(\mathbf{s}_t) \right \|^2
	\end{equation}
	where $V(\mathbf{s}_t) $ is the target value, $V_\phi(\mathbf{s}_t)$ is the value predicted by the critic network with its parameter set $\phi$. 
	Eventually, the training for comment generation is completed after $\mathcal{L}(\phi)$ converges.
	
	Denoting all the parameters as $\Theta=\{\theta, \phi \}$, the total loss of our model can be represented as $\mathcal{L}(\Theta)=\mathcal{L}(\theta)+ \mathcal{L}(\phi)$.
	We employ stochastic gradient descend with the diagonal variant of AdaGrad \cite{duchi2011adaptive} to tune the parameters of \approach{} for optimizing the code summarization model.

	\subsection{Code Search}
	
	Given a natural language query, \approach{} encodes all the code snippets and the generated comments into vector sets by tree-transformer-based encoder and transformer-based encoder respectively, and encodes the query into a vector by transformer-based encoder. Next, we compute the similarity scores between the query vector and the vectors in both the code snippets vector set and the generated comments vector set. 
	At last, we rank all the code snippets to recommend the search results derived from the linear combination of the two similarity score sets which are trained for optimality. 
	
	As shown in Figure \ref{fig:overview} (b), we encode the code snippets and the generated comments into vector spaces $\{V_c\} $ and $\{V_s\}$ by the tree-transformer-based encoder and transformer-based encoder respectively. We also encode the given natural language query into a vector $V_q$ by transformer-based encoder. Next, we compute the similarity scores between $V_q$ and the vectors of the code snippet $i$ from both $\{V_c\} $ and $\{V_s\}$ as $sim(V_q, V_{ci})$ and $ sim(V_q, V_{si})$.	
	Furthermore, we derive the weighted score of the code snippet $i$, $score_i$, by linearly combining $sim(V_q, V_{ci})$ and $ sim(V_q, V_{si})$, as shown in Equation \ref{eq:score}. Eventually, we rank all the code snippets according to their $score$s for recommending the search results,
	
	\begin{equation}\label{eq:score}
	score(Q, C) = \beta * sim(V_q, V_{ci}) + (1-\beta) * sim(V_q, V_{si})
	\end{equation}
	
	where $ \beta $ is a parameter that ranges from 0 to 1 and determined after training, $sim()$ is computed by consine. Specially, given the query $q_i$, the training objective is to ensure $score(q_i, c_i) > score(q_i, c_j)$, where $c_j$ demonstrates the code snippets in the dataset expect $ c_i$.

\section{Evaluation}\label{sec:experiment}
	We conduct a set of extensive experiments on the effectiveness and efficiency of \emph{\approach{}} in terms of both the code summarization and code search components compared with state-of-the-art approaches.	
	
	\subsection{Experimental Setups}\label{sec_dataset}
	To evaluate the performance of our proposed approach, we use the dataset presented in \cite{Barone2017A} where over 120,000 $<$code;comment$>$ pairs are collected from various Python projects in GitHub \cite{github} with 50,400 code tokens and 31,350 comment tokens in its vocabulary respectively. For cross validation, we shuffle the original dataset and use the first 60\% for training, 20\% for validation, and the rest for testing. 
	
	In our experiments, the word embedding size is set to 1280, the batch size is set to 2048, the layer size is set to 6, and the head number is set to 8. We pretrain both actor network and critic network with 20000 steps each, and train the actor-critic network with 100000 steps simultaneously. For the code search part, the comments in the dataset are utilized for the query. All the experiments in this paper are implemented with Python 3.5, and run on a computer with a 2.8 GHz Intel Core i7 CPU, 64 GB 1600 MHz DDR3 RAM, and a Saturn XT GPU with 24 GB memory running RHEL 7.5.

	\subsection{Result Analysis}\label{sec_result} 
	
	\subsubsection{Code summarization}
	To evaluate the code summarization component of \approach{}, we select several state-of-the-art approaches, i.e., Hybrid-DeepCom \cite{hu2019deep}, CoaCor \cite{yao2019coacor}, and AutoSum \cite{wan2018improving} for performance comparison with \approach{}. In particular, Hybrid-DeepCom \cite{hu2019deep} 
	utilizes the AST sequence converted by traversing the AST as the code representation and input the AST sequence to the GRU-based NMT for code summarization via combining lexical and structure information.
	CoaCor \cite{yao2019coacor} utilizes the plain text of source code and an LSTM-based encoder-decoder framework for code summarization.
	AutoSum \cite{wan2018improving} utilizes a tree-LSTM-based NMT model and inputs the code snippet as plain text to the code-generation model with reinforcement learning for performance enhancement. Similarly, the evaluation for \approach{} is also designed to explore the performance of its different components, where \approach{}$_{base}$ adopts the transformer-based encoder; \approach{}$_{tree}$ adopts the tree-transformer-based encoder for source code; and \approach{}$_{tree+RL}$, i.e., the complete \approach{}, utilizes the tree-transformer-based encoder for source code and reinforcement learning for further enhancing the code summarization model.

	We evaluate the performance of all the approaches based on four widely-used evaluation metrics adopted in neural machine translation and image captioning: BLEU \cite{papineni2002bleu}, METEOR \cite{banerjee2005meteor}, ROUGE \cite{lin2004rouge} and CIDER \cite{vedantam2015cider}. In particular, BLEU measures the average n-gram precision on a set of reference sentences with a penalty for short sentences. METEOR evaluates how well the results capture content from the references via recall which is computed via stemming and synonymy matching. 
	ROUGE-L imports account sentence level structure similarity and identifies the longest co-occurrence in sequential n-grams. CIDER is a consensus-based evaluation protocol for image captioning that evaluates the similarity of the generated comments and the ground truth. 
	
		\begin{table}[!t]
		\centering
		\caption{Code summarization results with different metrics. (Best scores are in boldface.)}  
		\label{tab:summarization}
		\renewcommand\arraystretch{1.6}
		\begin{tabular}{|p{2cm}||c|c|c|c|}
			\hline
			Approaches & \textbf{BLEU-1} & \textbf{METEOR} & \textbf{ROUGE-L} & \textbf{CIDER} \\
			\hline
			\hline
			Hybrid-DeepCom      & 15.60		& 6.09	&14.33	&51.88         \\
			\cline{1-5}
			CoaCor      & 25.60		& 9.52	&29.38	& 78.11         \\
			\cline{1-5}
			AutoSum     & 25.27		&9.29	&39.13	&75.01         \\
			\cline{1-5}
			\approach{}$_{base}$      & 	27.69	& 10.26	& 41.87	& 81.01         \\
			\cline{1-5}
			\approach{}$_{tree}$      & 32.05		&11.74	& 45.92	& 84.56        \\
			\cline{1-5}
			\approach{}$_{tree+RL}$  & \textbf{37.69}	& \textbf{13.52}	&\textbf{51.27}	& \textbf{87.24}       \\
			\hline    
		\end{tabular}
	\end{table}

		\begin{table*}[!t]
			\centering
			\caption{Sample issues for code summarization case study}
			\label{tab:study}
			\renewcommand\arraystretch{1.5}
			\begin{tabular}{|p{0.4cm}||p{3.3cm}|p{3cm}|p{9cm}|}
				\hline
				 & \textbf{Issue link} & \begin{tabular}{@{}l@{}l@{}}\textbf{Generated}\\ \textbf{comment}\end{tabular} & \textbf{Feedback} \\
				\hline
				\hline
				1 &	\url{https://github.com/mikunit567/GAE/issues/1}	 & Validate a given xsrf token by retrieving it.	& \textit{``Yes, this is correct. Validate a retrieved XSRF from the memory cache and then with the token perform an associated action.''}          \\
				\cline{1-4}
				 2 & \url{https://github.com/hamzafaisaljarral/scoop/issues/1}	& Iterates through the glob nodes.	& \textit{``Yup you have got that right but for better understanding you have to look into django-shop documentation and look into django-cms documentation as well.''}         \\
				\cline{1-4}
				 3 & \url{https://github.com/rumd3x/PSP-POC/issues/1}	& Combine two lists in a list. 	& \textit{``The \texttt{pstats} package is used for creating reports from data generated by the Profiles class. The \texttt{add\_callers} function is supposed to take a \texttt{source} list, and a \texttt{target} list, and return \texttt{new\_callers} by combining the call results of both target and source by adding the call time.''}         \\
				\hline
			\end{tabular}
		\end{table*}

	Table \ref{tab:summarization} demonstrates the code summarization results of all the approaches in terms of the selected metrics. While the compared approaches achieve close performances, e.g., around 20\% in terms of BLEU-1, \approach{} can approximate 38\%.  
	In particular, we can obtain the following detailed findings. First, we can observe that \approach{} can significantly outperform all the compared approaches in terms of all the evaluated metrics. For instance, the complete \approach{}, i.e., \approach{}$_{tree+RL}$ can outperform all the compared approaches from 47.2\% to 141.6\% in terms of BLEU-1. Such performance advantages can indicate the superiority of the transformer-enabled self-attention mechanism over the mechanisms, including the attention mechanism, that are adopted in other RNN-based approaches, because the self-attention mechanism can effectively capture the impacts of the overall text on all the tokens of the input sequences for better reflecting their semantics and thus optimizing the language model weights. Next, we can verify that each component of \approach{} is effective for enhancing the performance. For instance, by applying the tree-transformer-based encoder, \approach{}$_{tree}$ can dramatically outperform \approach{}$_{base}$ that only applies the transformer-based encoder by 15.7\% in terms of BLEU-1. We can verify that our tree transformer based on identifying and leveraging the indent-based program semantic granularity can effectively strengthen the language model by augmenting the semantic level information for tokens. Moreover, by applying reinforcement learning, \approach{}$_{tree+RL}$ outperforms \approach{}, i.e., \approach{}$_{tree}$ by 17.5\% in terms of BLEU-1, which can further verify the strength of reinforcement learning as verified in \cite{yao2019coacor, wan2018improving}.  Note that the performance of certain approaches, e.g., Hybrid-DeepCom,  dramatically differs from its original performance in \cite{hu2019deep} mainly because of the training data and programming language differences.

	We also conduct a set of case studies to further evaluate the effectiveness of \approach{}. In particular, we first collect Python projects from GitHub and input them to our \approach{}-trained model for generating their corresponding comments. Next, we issue such generated comments to the corresponding developers for their evaluations on the quality of the generated comments. In total, we received \response{} responses, among which \rights{} developers confirmed the correctness of the generated comments to summarize their code snippets. In addition, \explain{} developers extended detailed explanations of the associated code which also expose their support to our generated comments. The rest responses are unrelated to the correctness of our generated comments.   
	Table \ref{tab:study} presents selected examples of the developer feedback where the first and second case indicate that the developers confirm the correctness of our generated comments while the third case reveals that the developer is supportive to the generated comment though he did not directly present it. 

\subsubsection{Code search}
    To evaluate the effectiveness of the code search component of \approach{}, we select several state-of-the-art approaches for comparison. Firstly, for the aforementioned approaches Hybrid-DeepCom, AutoSum, and CoaCor, we utilize the generated comments of those approaches as the input for the code search part of \approach{}. In addition to further utilizing them for code search, we also compare \approach{} with DeepCS \cite{gu2018deep} which utilizes RNN to encode code and query and compute the distance between the code vector and the query vector for returning the code snippets with the closest vectors.
	The performance of code search is evaluated in terms of four widely-used metrics: MRR (Mean Reciprocal Rank) \cite{Craswell2009Mean}, nDCG (normalized Discounted Cumulative Gain) \cite{Wang2013A} and Success Rate@k \cite{Xuan2016Relationship}, where MRR measures the average reciprocal ranks of results given a set of queries and the reciprocal rank of a query is computed as the inverse of the rank of the first hit result; 
	nDCG considers the ranking of the search results which evaluates the usefulness of result based on its position in the result list; and Success Rate@k measures the percentage of queries for which more than one correct result exist in the top $k$ ranked results.

\begin{table}[!t]
		\centering
		\caption{Code search accuracy compared with baselines. (Best scores are in boldface.)}
		\label{tab:search}
		\renewcommand\arraystretch{1.6}
		\begin{tabular}{|p{2.4cm}||c|c|c|c|c|}
			\hline
			Approaches & \textbf{MRR}  & \textbf{nDCG} & \textbf{SR@5} & \textbf{SR@10}  \\
			\hline
			\hline
			DeepCS      & 48.41 	& 58.85 & 57.44 & 66.78	         \\
			\cline{1-5}
			CoaCor      & 59.33	  & 67.51 & \textbf{67.05} & 73.58	         \\
			\cline{1-5}
			Hybrid-DeepCom		& 50.92	& 59.92  & 60.52 & 68.35	        \\
			\cline{1-5}
			AutoSum     & 57.68		& 63.52  & 63.43 & 70.16	         \\
			\cline{1-5}
			\approach{}$_{base}$      & 58.43	& 65.13  & 63.28 & 70.85         \\
			\cline{1-5}			 
			\approach{}$_{tree}$      & 60.57	& 68.43	 & 65.16 & 74.13         \\
			\cline{1-5}
			\approach{}$_{tree+RL}$  & \textbf{62.37}	&\textbf{70.62}	  & 66.95 & \textbf{75.21}       \\
			\hline    
		\end{tabular}
	\end{table}
	
	Table \ref{tab:search} shows the code search result comparisons between our proposed approach and the aforementioned baselines where we can observe that \approach{} can outperform all the other approaches in terms of all the evaluate metrics. Specifically, in terms of MRR, \approach{}$_{tree+RL}$ can outperform all the other approaches from 5.12\% to 28.8\%. Compared with the code summarization results, the advantages of \approach{} over the same adopted approaches on code search dramatically shrinks which can be discussed as follows: (1) the code search metrics are naturally subject to less distinguishable results than the code summarization metrics. For Hybrid-DeepCom, AutoSum, and \approach{} which all utilize the generated comments to strengthen their code search performance, their adopted code summarization metrics are essentially based on word frequency which generally are fine-grained, e.g., BLEU-based metrics, while their code search metrics are generally based on coarse-grained query-wise comparisons. Therefore, the code summarization metrics tend to result in distinguishable results for different techniques because they are likely to reflect the trivial difference between two generated comments. However, their corresponding code search results might not be that distinguishable because the two generated comments might be trained to result in the result in the identical code rankings. For instance, suppose two code summarization approaches generate the comments ``\textit{returns the path of the target dependencies}'' and ``\textit{derive a target-dependency list}'' respectively. While they can be used to represent the same code snippets, they may result in different BLEU scores because they consist of different words. However, if they are used for code search, they can both rank the code snippet of Figure 2 on the top and thus result in the identical score in terms of the code search metrics. 
	(2) CoaCor \cite{yao2019coacor} can approach a close performance to \approach{} because its rewarding mechanism utilizes the search accuracy to guide the code annotation generation and search modeling directly. 
	However, We can observe that \approach{} significantly outperforms CoaCor in terms of code summarization (by 47.2\%). Therefore, to bridge such performance gap, CoaCor has to pay extra effort for enhancing its modeling process while \approach{} can limit its effort in training the model once and for all for optimizing both code summarization and code search.

	We also conduct a case study to evaluate the effectiveness of \approach{}. We organized 5 postgraduate students and 5 developers with certain Python background. 
	We designed 15 programming tasks where each participant is asked to choose 3 tasks for code search using \approach{} as well as our benchmark. Two example tasks are listed as follows:
	\begin{itemize}
	\item
	Task 1: Remove all the files in a directory.
	\item
	Task 2: Sends a message to the admins.
	\end{itemize}
	
	Then, they are asked to evaluate if the searched code snippets can solve the tasks or are helpful for solving them, by giving a score on a five-point Likert scale (strongly agree is 5 and strongly disagree is 1). For the 10 participants, the average Likert score is 3.167 (with standard deviation of 1.472), which indicates that in general, the efficacy of \approach{} can be acceptable.

\section{Threats to Validity}\label{sec:threats}
There are several threats to validity of our proposed approach and its results, which are presented as follows. 

The main threat to internal validity is the potential defects in the implementation of our techniques. To reduce such threat, we adopted a commonly-used benchmark with over 120,000 Python functions for evaluating the effectiveness and efficiency of our proposed approach and several existing approaches for comparison. Moreover, to ensure the fair comparison, we directly downloaded the optimized models of the existing approaches for comparison.

The threats to external validity mainly lie in the dataset quality and the evaluation metrics of our experiments. On one hand, the quality of the training data, i.e., the $ <code; comment> $ pairs adopted in our experiment was not evaluated. Among the over 120,000 python functions, it is likely that part of the poor-quality data can taint the training results. However, since (1) all the approaches were evaluated in the identical benchmark, and (2) the adopted evaluation metrics measure the performance of the approaches by word frequency where the corresponding performance difference among the approaches can indicate their word mapping levels, we can also infer that given high-quality training data, the performance distribution of all the approaches are likely to maintain consistency, where \approach{} can still outperform the other approaches in terms of the word-frequency metrics. Moreover, the performance of the tree-transformer-based encoder heavily relies on the quality of program forms. However, the experimental results indicate that the tree-transformer-based encoder can achieve better performance than the transformer-based encoder regardless the quality of the program forms. On the other hand, the word-frequency-based metrics cannot fully reflect the the semantic correctness of the approaches. To reduce such threat, we adopted a set of empirical studies such that developers can feedback for the quality of our code summarization and code search results. The positive study results can strengthen the validity of the effectiveness and efficiency of \approach{}.
	
\section{Related Work}\label{sec:relatedwork}

	\subsection{Code Summarization}
    The code summarization techniques can be mainly categorized as information-retrieval-based approaches and deep-learning-based approaches.

    \textbf{Information-retrieval-based approaches. } 
    Wong et al. \cite{wong2013autocomment} proposed AutoComment which leverages code-description mappings to generate description comments for similar code segments matched in open-source projects. Similarly they also apply code clone detection techniques to find similar code snippets and extract comments from the similar code snippets \cite{wong2015clocom}. 
    Movshovitz-Attias et al. \cite{movshovitz2013natural} predicted comments from Java source files using topic models and n-grams.
    Haiduc et al. \cite{Haiduc2010On} combined IR techniques, i.e., Vector Space Model and Latent Semantic Indexing, to generate terms-based summaries for Jave classes and methods.

\textbf{Deep-learning-based approaches. }
The deep-learning-based approaches usually leverage Recurrent Neural Networks (RNNs) or Convolution neural networks (CNNs) with the attention mechanism. For instance, Iyer et al. \cite{iyer2016summarizing} proposed to use RNN with an attention mechanism---CODE-NN to produce comments for C\# code snippets and SQL queries. Allamanis et al. \cite{allamanis2016convolutional} proposed an attentional CNN on the input tokens to detect local time-invariant and long-range topical attention features to summarize code snippets into function name-like summaries.
Considering the API information, Hu et al. \cite{hu2018summarizing} proposed TL-CodeSum to generate summaries by capturing semantics from the source code with the assistance of API knowledge.
Chen et al. \cite{chen2018neural} proposed BVAE which utilizes C-VAE to encode code and L-VAE to encode natural language. 
In addition to such encoder-decoder-based approaches, Wan et al. \cite{wan2018improving, drlcomment} drew on the insights of deep reinforcement learning to alleviate the exposure bias issue by integrating exploration and exploitation into the whole framework. 
Hu et al. \cite{hu2018deep} proposed DeepCom which takes AST sequence converted by traversing the AST as the input of NMT and they also extended this work by considering hybrid lexical and syntactical information in \cite{hu2019deep}.
Leclair et al. \cite{leclair2019neural} combined words from code with code structure from AST, which allows the model to learn code structure independent of the text in code.
	
	\subsection{Code Search}
Code search techniques also mainly consists of information-retrieval-based approaches and deep-learning-based approaches.
	
	\textbf{Information-retrieval-based approaches. } 
	Hill et al. \cite{Hill2014NL} proposed CONQUER which integrates multiple feedback mechanisms into the search results view. 
	Some approaches proposed to extend the queries, for example, 
	Lu et al. \cite{lu2015query} proposed to extend queries with synonyms generated from WordNet and then match them with phrases extracting from code identifiers to obtain the search results.
	Lv et al. \cite{lv2015codehow} designed a API understanding component to figure out the potential APIs and then expand the query with the potential APIs and retrieve relevant code snippets from the codebase. Similarly, Raghothaman et al.  \cite{raghothaman2016swim} proposed \emph{swim}, which first suggests an API set given a query by the natural language to API mapper that is extracted from clickthrough data in search engine, and then generates code using the suggested APIs by the synthesizer.

	\textbf{Deep-learning-based approaches}
	The deep learning-based approaches usually encode the code snippets and natural language query into a hidden vector space, and then train a model to make the corresponding code and query vector more similar in the hidden space. 
	Gu et al. \cite{gu2018deep} proposed DeepCS, which reads code snippets and embeds them into vectors. Then, given a query, it returns the code snippets with the nearest vectors to the query. 		
	Luan et al. \cite{luan2018aroma} proposed Aroma, which takes a code snippet as input, assembles a list of method bodies that contain the snippet, clusters and intersects those method bodies to offer code recommendations.
	Different from the above approaches, Akbar et al. \cite{akbar2019scor} presented a framework that incorporates both ordering and semantic relationships between the terms and builds one-hot encoding model to rank the retrieval results. 	
	Chen et al. \cite{chen2018neural} proposed BVAE, which includes C-VAE and L-VAE to encode code and query respectively, based on which semantic vector for both code and description and generate completely. 	
	Yao et al. \cite{yao2019coacor} proposed CoaCor, which designs a rewarding mechanism to guide the code annotation model directly based on how effectively the generated annotation distinguishes the code snippet for code retrieval. 
	
	\textbf{Other approaches.} Sivaraman et al. \cite{sivaraman2019active} proposed ALICE, which integrates active learning and inductive logic programming to incorporate partial user feedback and refine code search patterns.  Takuya et al. \cite{takuya2011spontaneous} proposed Selene, which uses the entire editing code as query and recommends code based on a associative search engine.  Lemons et al. \cite{Lemos2007CodeGenie} proposed a test-driven code search and reuse approach, which searches code according to the behavior of the desired feature to be searched.

\section{Conclusion}\label{sec:conclusion}
	In this paper, we propose \approach{}, which is a transformer-based framework to integrate code summarization with code search. Specifically, \approach{} enables an actor-critic network for code summarization. In the actor network, we build transformer- and tree-transformer-based encoder to encode code snippets and decode the given code snippet to generate their comments. Meanwhile, we utilize the feedback from the critic network to iteratively tune the actor network for enhancing the quality of the generated comments. Furthermore, we import the generated comments to code search for enhancing its accuracy. We conduct a set of experimental studies and case studies to evaluate the effectiveness of \approach{}, where the experimental results suggest that \approach{} can significantly outperform multiple state-of-the-art approaches in both code summarization and code search and the study results further strengthen the efficacy of \approach{} from the developers' points of view.


\bibliographystyle{IEEEtran}
\bibliography{ref}

\appendix

\end{document}